\documentclass[preprint,12pt]{elsarticle}
\usepackage{float}
\usepackage{here}

\usepackage{capt-of}
\usepackage{subfig} 
\usepackage{hyperref}
\usepackage{chngcntr}
\counterwithout{figure}{section}
\usepackage{amsmath}
\usepackage{lipsum}

\begin{document}

\title{Numerical Investigation of Water Entry of Hydrophobic Spheres.}  

\author[mydepartment]{Jaspreet Singh}
\author[mydepartment]{Anikesh Pal\corref{mycorrespondingauthor}}
\cortext[mycorrespondingauthor]{Corresponding author}
\ead{pala@iitk.ac.in}

\address[mydepartment]{Department of Mechanical Engineering, Indian Institute of Technology Kanpur, India}

\begin{abstract}
We perform numerical simulations to study the dynamics of the entry of hydrophobic spheres in a pool of water using ANSYS. To track the air-water interface during the translation of the sphere in the pool of water, we use the volume of fluid (VOF) model. The continuum surface force (CSF) method computes the surface tension force. To simulate the hydrophobic surface properties, we also include wall adhesion. We perform simulations with different diameters and impact speeds of the sphere. Our simulations capture the formation of different types of air cavities, pinch-offs of these cavities, and other finer details similar to the experiments performed at the same parameters. Finally, we compare the coefficient of drag among the different hydrophobic cases. We further perform simulations of hydrophilic spheres impacting the pool of water and compare the drag coefficient with the analogous hydrophobic cases. We conclude that the spheres with hydrophobic surfaces have a lower drag coefficient than their hydrophilic counterparts. This lower drag of the hydrophobic spheres is attributed to the formation of the air cavity by the hydrophobic surfaces while translating through the pool of water, which reduces the area of the sphere in contact with water. In contrast, no such air cavity forms in the case of hydrophilic spheres.\\
\end{abstract}

\maketitle 

\section{Introduction}

The vertical impact of a solid object with water has been of interest to many researchers owing to the formation of an air cavity and the subsequent translation of the object in the water. This entire process plays a significant role in a variety of problems. In military applications \citep{birkhoff1949fluid,birkhoff1951transient,birkhoff1957jets,may1951effect,may1952vertical,may1975water} understanding the water entry phenomenon can assist in designing air-to-sea launchable projectiles that can preciously neutralize the underwater sea mines and torpedoes. Similarly, in marine applications \citep{asfar1987rigid, Cointe,karagiozva1996dynamic}, the intermittent entry of the ships or ocean structures during harsh conditions results in impact loading that can cause damage to ship hulls and ocean structures. The knowledge of the water entry phenomenon can help in estimating the impact forces and can subsequently assist in predicting the failure of marine structures. Apart from these engineering applications, water entry of solid objects find applications in bio-locomotion \citep{bush2006walking} and astrophysical phenomenon \citep{melosh1999impact,thoroddsen2001granular}. \\

The phenomenon of water entry of the sphere has been characterized by the Weber number $We= \rho W^2R_0/\sigma$, the Bond number $Bo=\rho gR_0^2/\sigma$, and the Reynolds number $Re=\rho W_0R_0/\mu$. The Weber number represents the relative magnitude of the inertia and surface tension forces, while the Bond number describes the relative magnitude of the gravitational forces to the capillary forces. Typically the Froude number, $Fr=W_0^2/gR_0=We/B$, has been used as a parameter to define different regimes of the water entry phenomenon based on the type of cavity collapse. \\ 

The earlier experiments reporting the qualitative and quantitative characteristics of the physical events associated with the various stages of the water entry of a sphere were performed by \cite{worthington1897v,gilbarg1948influence}. Theoretical investigations were carried out by \cite{korobkin1988initial, howison1991incompressible, oliver_2007} to study the normal impact of a body with a liquid surface. \cite{thoroddsen2004impact} experimentally studied the initial stage associated with the impact of a solid sphere onto a liquid surface. They reported that at a high Reynolds number, a high-speed horizontal jet emerges immediately after the initial contact of the sphere with water. \\

The initial stage of impact is followed by the formation, pinch-off, and growth of air cavities. \cite{glasheen1996vertical} and \cite{gaudet1998numerical} carried out experimental and numerical investigations of water entry of circular disks at different $Fr$ to study these follow up phenomena. \cite{gekle2008noncontinuous} studied the cavity formation by pulling the cylinder vertically through a water surface at a constant speed and demonstrated that the resulting air cavity collapses owing to the hydrostatic pressure, leading to a rapid and axisymmetric pinch-off in a single point. They also found two separate scaling regimes of pinch-off depths with varying $Fr$ and correlated all these observations with the capillary wave effect. \cite{bergmann2006giant} experimentally studied the dynamics of a giant cylindrical air cavity formed owing to the translation of a disk through the water. They found that for finite values of $Fr$. the collapse of the air-cavity is not a strictly self-similar phenomenon. However, in the limit of infinite $Fr$, the collapse of the air-cavity is self-similar. The formation and collapse of an air cavity induced by high-speed impact and penetration of a rigid projectile into the water were analytically investigated by \cite{lee1997cavity}. It was deduced that the time of pinch-off of the air cavity is independent of the diameter of the sphere at high impact speeds, whereas the location of the pinch-off is weakly dependent on the impact velocity. A theoretical approach was also adopted by \cite{duclaux2007dynamics} to study the complete process of formation to collapse of a transient cavity of air in the water created by the impact of a cylinder and sphere at high $Re$ and high $We$. Their analytical solution concludes that the cavity dynamics for a cylindrical and a spherical object are very different and depend on $Fr$. \\

The static water contact angle $\theta$ of a sphere is an important parameter governing the dynamics of water entry. When $\theta \ge 90^{\circ}$ the surface of the sphere is not wettable, and the sphere is hydrophobic, whereas if $0^{\circ} \le \theta \le 90^{\circ}$ the surface of the sphere is wettable and the sphere is called hydrophilic. \cite{duez2007making} concluded that the degree of splash made by an object during water entry depends on the wettability of the object. \cite{aristoff_bush_2009} performed experiments and theoretical modeling of water entry of a small hydrophobic sphere of $R_0 << 2.7 mm$ at low $Bo$. They reported that the air cavity formed during the impact and the translation through water collapses owing to the combined effect of surface tension and gravity. Their parametric study revealed the formation of four distinct types of cavities, namely quasi-static, shallow seal, deep seal, and surface seal, with the increase in $We$. In a quasi-static cavity, the air entrainment is minimal, and the cavity takes the shape of a hydrostatic meniscus. A considerable quantity of air is entrained in both the shallow seal and deep seal cavities, resulting in the formation of a long slender cavity. Capillary waves also generate during the formation of such cavities. These two types of cavities differ in terms of their pinch-off locations. The surface seal cavity is characterized by the closure of the splash curtain owing to some combination of the curvature pressures and aerodynamic pressures before its pinch off at depth. \\

\cite{jiang2016numerical} performed numerical simulations to study the transient behavior of water-entry supercavitating flow around a cylindrical projectile at different impact velocities in the presence of turbulent drag-reducing additives. They compared the cavity lengths, velocity attenuations, penetration distances, and drag coefficients for water and drag-reducing solution cases at high and low impact velocity cases. They reported a significant enhancement in supercavitation and reduction in drag owing to the drag-reducing additives at low velocities compared to high velocities. Numerical simulations of spheres made of different materials entering water at an initial speed of $2.17$ m/s were carried out by \cite{ahmadzadeh2014numerical} using an Eulerian-Lagrangian method, and the results were verified against the experiments.\\

\begin{figure}[ht]
    \centering
   (a) \includegraphics[width=8cm]{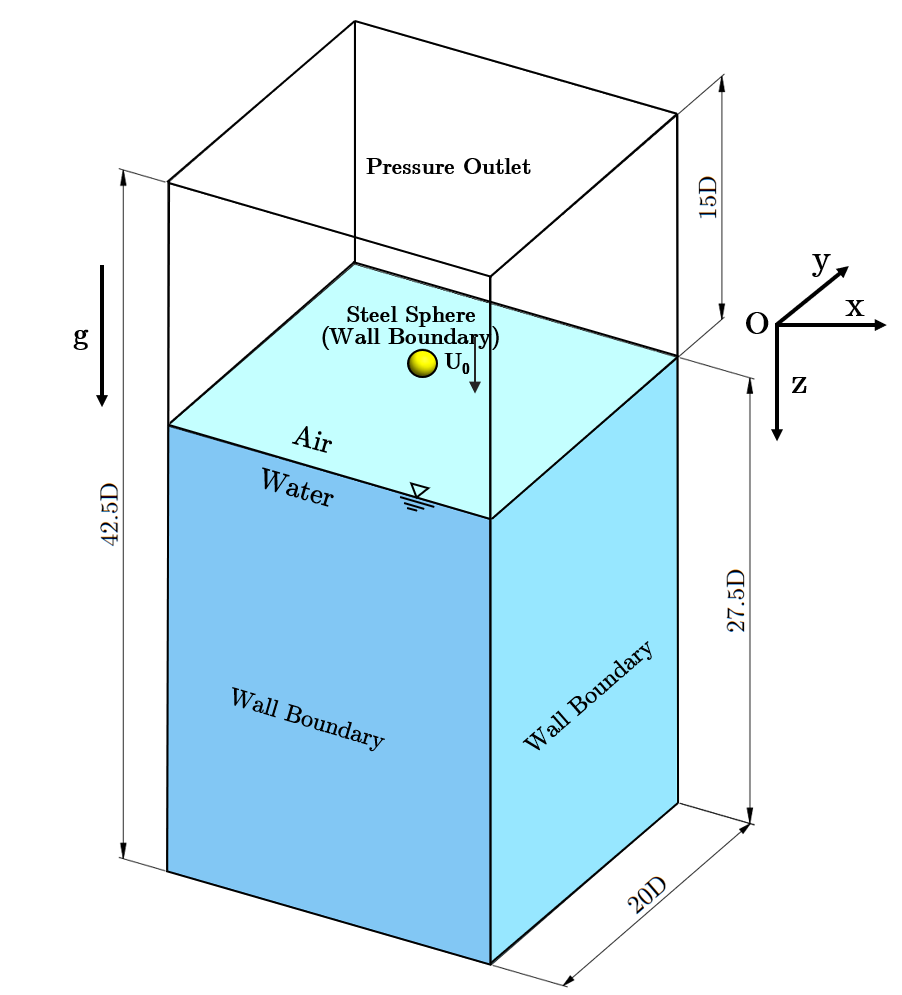}
   (b)\includegraphics[width=5.5cm]{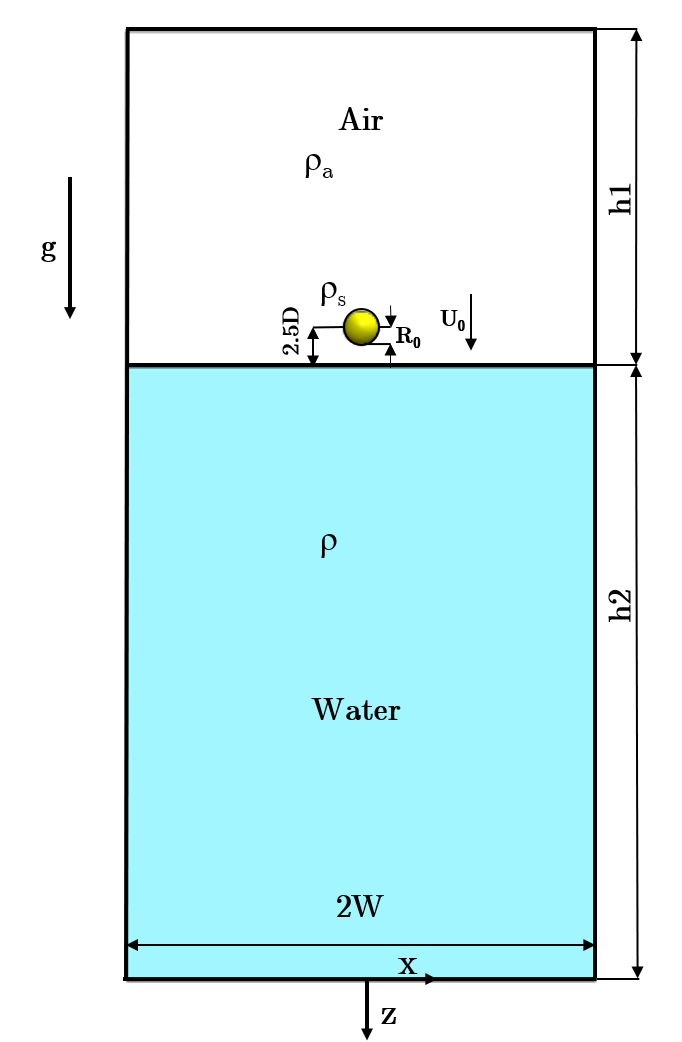}
    \caption{Computational domain for the simulations of a solid sphere entering a pool of water. The sphere of radius $R_0$ (diameter D) impacts the water pool with velocity $W_0$. The sphere's center is at the origin. The half-width of the domain is 10D, the height above the water surface h1 is 13D+2mm, and the depth of the water from surface h2 is 27.5D in all the simulations. In the case of the sphere with impact velocity $5.4m/s (We=420, B=0.14)$, we took h2= 65D. \textbf{Boundary conditions of the numerical model:} Top of the tank is the pressure outlet, and the pressure is set up at atmospheric default condition. We use wall boundary conditions for the side walls and the bottom of the tank. The domain is symmetric about the $y$-axis. No-slip wall boundary condition is used for the surface of the solid sphere, whereas we use free slip condition at the side walls.}
\label{fig:1}
\end{figure}

In the present study, we perform three-dimensional numerical simulations to explore the phenomenon of water entry and the subsequent translation of a hydrophobic sphere in water. We consider a hydrophobic solid sphere of radius $R_0$, and density $\rho_s$ moving in the downward direction with a velocity of $W_0$ and impacting a fluid of density $\rho$, viscosity $\mu$ and surface tension of $\sigma$. The numerical setup is in figure {\ref{fig:1}}. We use a volume of fluid function to represent the air-water interface. Our focus is to capture various types of air cavities and their evolution during the translation of the sphere through the water. Therefore, we present quantitative results for four different values of $W_0$ corresponding to four different $We$, similar to the experiments of \cite{aristoff_bush_2009}. We also simulate the analogous hydrophilic cases and present a comparison of the evolution of the drag force during water entry and translation for both hydrophobic and hydrophilic cases. \\

\section{Numerical Method}

We use \cite{AnsysFluent20} for mesh generation and the subsequent numerical calculations. The material and the surface properties of the solid sphere are kept constant in all the cases. The conservation of mass and momentum is represented by the following equations 

\begin{equation}
\label{mass}
\nabla \cdot \textbf{u}=0,
\end{equation}

\begin{equation}
\label{momentum}
\frac{\partial}{\partial t}(\rho \textbf{u}) + \nabla \cdot (\rho \textbf{u}\textbf{u}) = - \nabla p + \nabla \cdot [\mu(\nabla \textbf{u} + \nabla \textbf{u}^T )]+ \rho \textbf{g} + \textbf{F},
\end{equation}

where $\textbf{u}$ is the velocity vector, $p$ is the pressure, $\rho$ is the density, $\mu$ is the dynamic viscosity, $F$ is the surface tension force and $g$ is the acceleration because of gravity. The motion of the air-water interface is represented by
\begin{equation}
\label{VOF}
\frac{\partial{\alpha_l}}{\partial{t}} + \nabla \cdot (\textbf{u} \alpha_l) = 0,
\end{equation}
where 
$\alpha_l$ represents the liquid phase volume fraction \citep{brackbill1992continuum}. The density and viscosity are calculated using:
 \begin{equation}
 \rho= \rho_l\alpha_l+ \rho_g(1-\alpha_l),
 \end{equation}
 \begin{equation}
  \mu= \mu_l\alpha_l+\mu_g(1-\alpha_l).
 \end{equation}
Here $\rho_g$, $\rho_l$, are the densities, and $\mu_g$, $\mu_l$ are the viscosities of the gas and liquid phases, respectively. If the volume fraction of the fluid in the cell is $\alpha_g$(of the $g^{th}$ fluid ), then $\alpha_g$ =1, $\alpha_g$ =0, and 0<$\alpha_g$ <1 represents that cell is full of fluid, an empty cell, and the cell contains the interface between the fluid and other fluids respectively. In our cases $We$ and $Bo$ ranges from $1.9$ to $420$ and $0.088$ to $0.27$ respectively. We also include the gravitational forces along with the surface tension forces owing to their effect on the pinch-off depth of the cavity. \\

\begin{table}[ht]
    \centering 
    \begin{tabular}{@{}| c | c | c | c | c | c |@{}}
   \hline
        Case Velocity & Radius $(R_0)$ & $Re$ & $We$ & $Bo$ & Number of Mesh Elements \\\hline
        $0.3m/s$ & $1.4mm$ & $83.83$  & $1.9$ & $0.27$ & $1.25$ Million  \\ \hline
        $2.3m/s$ & $1mm$ & $4591.7$ & $72$ & $0.14$ & $1.25$ Million  \\ \hline
        $3.1m/s$ & $0.79mm$ & $4840.7$ & $109$ & $0.088$ & $1.1$ Million \\ \hline
        $5.4m/s$ & $1mm$ & $10780.5$ & $420$ & $0.14$ & $8$ Million  \\ \hline
    \end{tabular}
    \caption{Various parameters considered in the simulations}
    \label{Table:1}
\end{table}

\begin{table}[ht]
 \centering 
\begin{tabular}{ | c | c | c |} 
\hline
 Property &	Air phase &	Water phase \\ 
 \hline
 Density $kg/m^3$ & 1.225 & 998.2\\ \hline
 Dynamic viscosity $Pa$ $s$ & $1.7894e^{-5}$ & 0.001003 \\ \hline
 Surface Tension ($\sigma$) $N/m$ & - & 0.072\\ \hline
\end{tabular}
\caption{The properties of the water and air phase used in the simulations }
\label{Table:2}
\end{table}

We perform simulations for different sphere radii and impact velocities. The sphere density is $7700kg/m^3$ as the experiments of \cite{aristoff_bush_2009}. Since the radius and velocity of the sphere are changing, the values of $We$ and $Bo$ change accordingly. We list the different simulation parameters in table \ref{Table:1}. The contact angle $(\theta_c)$ of the sphere wall with water is $166^o$, thus, making the surface of the sphere super-hydrophobic. The contact angle for the hydrophilic cases is $14^o$. To reduce the computational cost, instead of simulating the entire dropping event of the sphere from the initial position with zero initial velocity, we positioned the sphere at $2$mm above the air-water interface with an approximate initial velocity in the normal direction of the water surface according to the velocity that it would be attained, based on the experiments \citep{aristoff_bush_2009}. For comparing the computational results with experimental findings, we calculate the reference time according to the initial position of the solid sphere in the experiments of \cite{aristoff_bush_2009}. The properties of the fluids used for the present simulations are in table \ref{Table:2}.\\

\cite{aristoff_bush_2009} ignored the presence of turbulent fluctuations in their theoretical derivations of the different pinch-off depth and time equations. Ignoring turbulent fluctuations is a reasonable assumption due to the formation of the air cavity. The air cavity prevents the formation of a wake of water past the sphere, making the flow highly streamlined around the sphere. Therefore, we assume laminar flow for the hydrophobic sphere for our simulations. However, for a hydrophilic sphere wake without an air cavity forms behind the sphere. Therefore, we use the $SST k-\omega$ turbulence model for simulating the hydrophilic case. Since the problem involves moving boundaries, we use three-dimensional dynamic mesh modeling to translate the sphere. \\

\section{Results and discussions}
\begin{figure}
\centering
    \includegraphics[width=10cm]{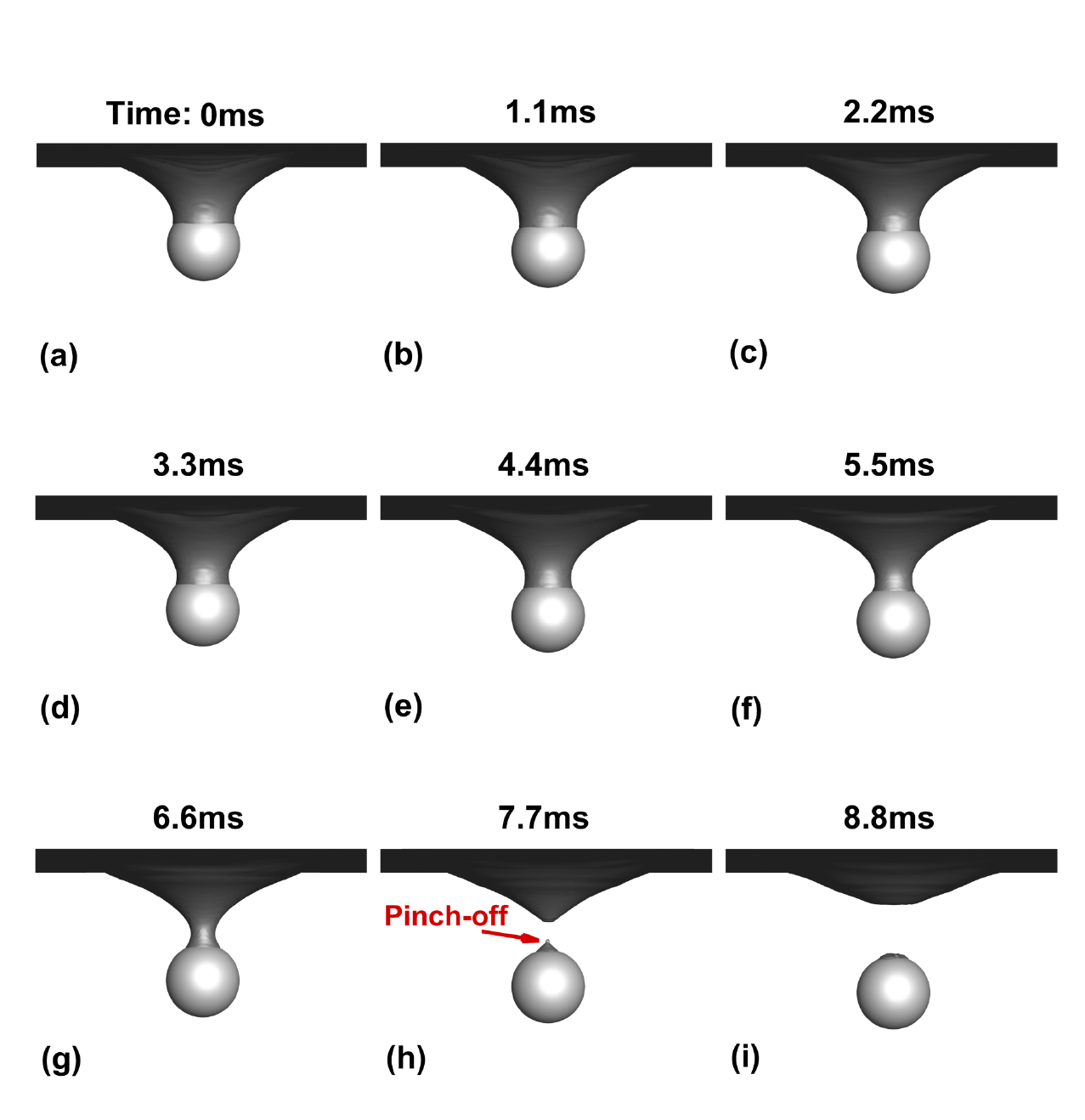}
    \caption{{Images of the water entry of a hydrophobic sphere with an impact speed $W_0= 0.3m/s$. The time interval between images is $1.1 ms; We = 1.9; B = 0.27.$}} %
    \label{fig:2}%
\end{figure}

When the hydrophobic sphere enters the pool of water, four different types of cavities form for low Bond numbers. The sphere impact on the air-water interface creates an axisymmetric air cavity that expands radially before closing under the combined influence of the hydrostatic pressure, the surface tension, and the aerodynamic pressure. For $B<<1$ and $1<We<<\tilde{D}^{-1}$ (density ratio of air-liquid $\rho_a/\rho$ ), the hydrostatic $(\rho g z)$ and the aerodynamic pressures are negligible as compared to the curvature pressures and pinch-off near the surface is expected as observed in the experiments of \cite{aristoff_bush_2009}. Figure \ref{fig:2} demonstrates the impact of a hydrophobic sphere at an impact velocity of $0.3m/s$, $We = 1.9$, and $B=0.27$. The sphere sinks and immerses in water after $t \sim 0.01 s$. The cavity collapses when the contact line (indicated by the arrow in figure \ref{fig:2}h) approaches the apex of the sphere at a depth of approximately the capillary length $\sqrt{\sigma/(\rho g)} =2.71mm$. The vertical and radial extents of the cavity are $5.45$ mm and $6.71$ mm, respectively, and are also of the order of the capillary length. We observe minimal air entrainment, and a tiny air bubble remains attached to the sphere after the pinch-off (see figure \ref{fig:2} h), similar to that observed in the experiments. This type of cavity is called a quasi-static cavity.\\

\begin{figure}
    \centering
    \includegraphics[width=18cm]{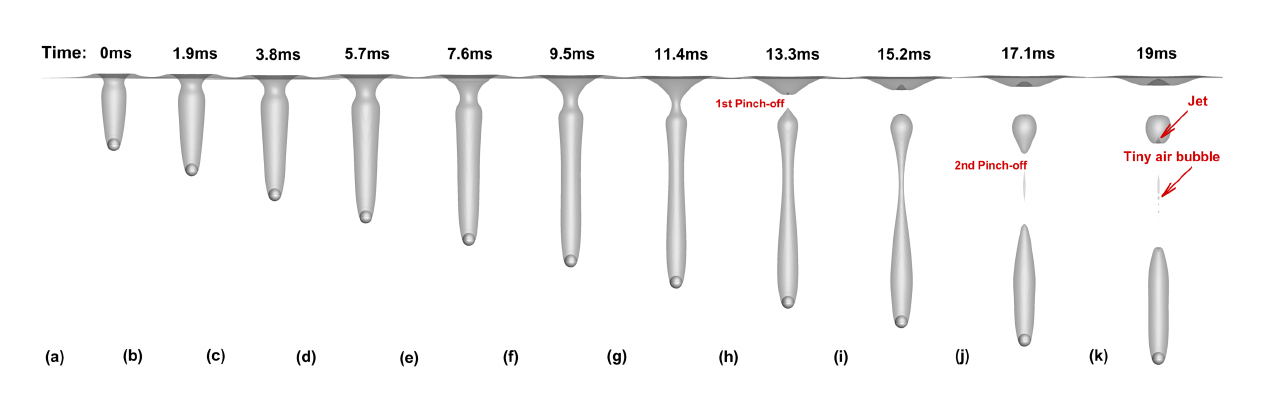}
    \caption{{Images of the water entry of a hydrophobic sphere with an impact speed $W_0= 2.3m/s$. The time interval between images is $1.9 ms; We = 72; B = 0.14.$}}
    \label{fig:3}
\end{figure}

Figure \ref{fig:3} demonstrates the translation of the sphere of radius $1$ mm after hitting the water surface with velocity $2.3$ m/s. A tongue-shaped cavity forms after the impact. As the sphere moves downwards the contact line remains fixed near the equator. The interface stretches upwards, resulting in a highly sloped cavity wall similar to the experiments. This cavity pinches off at a depth of $2.744$mm owing to the effect of surface tension force. This pinch-off depth is of the order of the capillary length and is similar to the first pinch-off depth of $2.714$mm found by \cite{aristoff_bush_2009}. This type of cavity is called shallow-seal due to its near-surface collapse. The volume of air entrained in this case is much greater than the size of the sphere compared to the previous case. This entertained volume of air further pinches off at a depth of $12.233$mm. Our simulation also captures the formation of the jet inside the secondary air bubble separated from the big air cavity, and the tiny air bubble (as indicated by the red arrows in figure \ref{fig:3}(k) between the air bubble separated from the big air cavity and air cavity attached to the sphere, similar to that observed in the experiment.\\

\begin{figure}[ht]
    \centering
    \includegraphics[width=16cm]{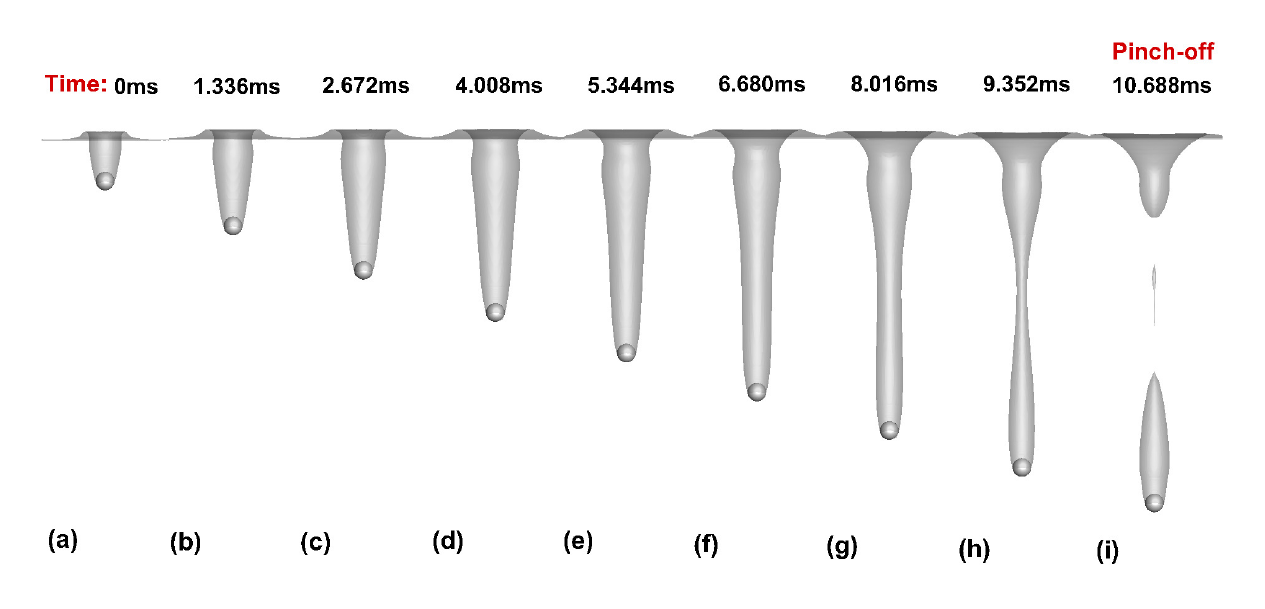}
    \caption{ {Images of the water entry of a hydrophobic sphere with an impact speed  $W_0= 3.1m/s$. The time interval between the images is $1.336ms; We = 109; Bo = 0.088$} }
    \label{fig:4}
\end{figure}

\begin{figure}[ht]
    \centering
    \includegraphics[width=14cm]{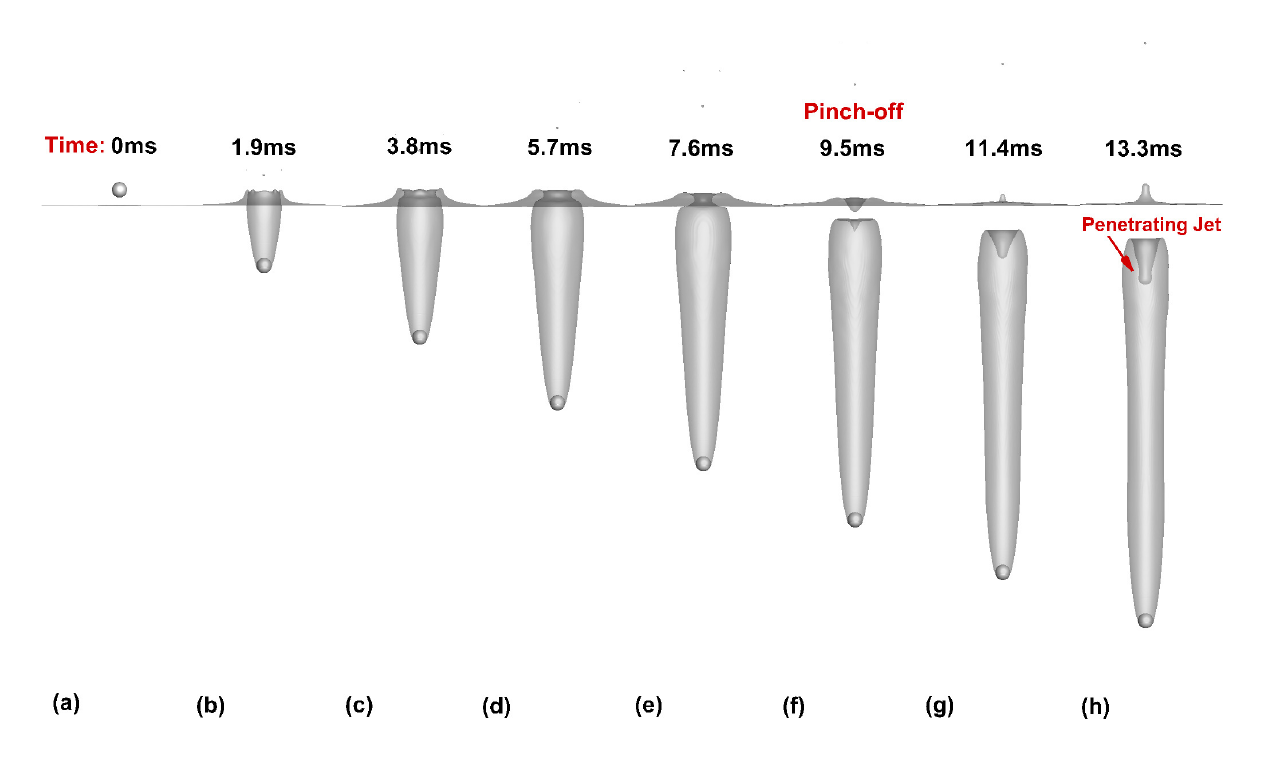}
    \caption{ {Images of the water entry of a hydrophobic sphere with an impact speed  $W_0= 5.4m/s$). The time interval between images is $1.9ms; We = 420; Bo = 0.14 $. }}
    \label{fig:5}
\end{figure}

Increasing the impact velocity of the sphere to $3.1$ m/s, but reducing the diameter to $0.79$ mm results in a cavity, as shown in figure \ref{fig:4}. A longer air cavity forms and the pinch-off occurs at a greater depth than the previous case. This type of cavity is called a deep seal cavity. The pinch-off location is approximately halfway between the sphere and the surface. The distance of the submerged sphere from the surface of the water when the pinch-off occurs is $31.568$mm and is similar to the experiment ($31.164$ mm).\\

Figure \ref{fig:5} shows the cavity dynamics for the sphere impacting water at $5.4$m/s. After the impact, a splash curtain forms that domes over to close the cavity from above. This collapse is due to the combined effect of the aerodynamic and curvature pressures acting on the splash curtain. Such a type of cavity is known as surface-seal. After the surface seal, the cavity expands, and the pressure inside it decreases, resulting in the detachment of the cavity from the surface. With an increase in the depth, a water jet penetrates the cavity from above owing to a larger hydrostatic pressure than the curvature pressure. Our simulation accurately captures this jet penetration phenomenon. The pinch-off depth for this case is $40.43$mm, which is close to that observed value of $38.11$mm in the experiments of \cite{aristoff_bush_2009}. \\

\begin{figure}[ht]
 \centering
 (a) \includegraphics[width=6cm]{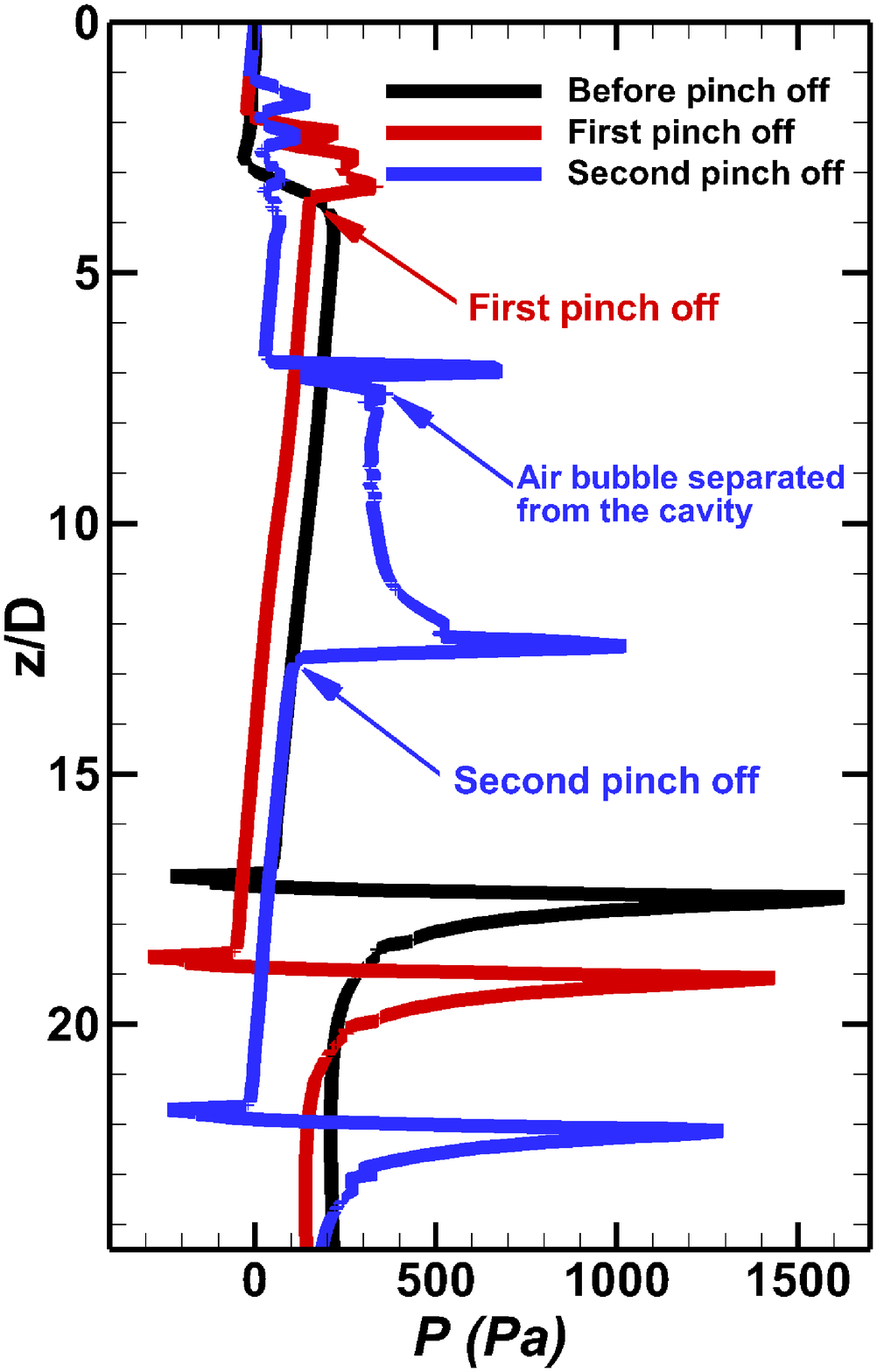}
 (b) \includegraphics[width=6cm]{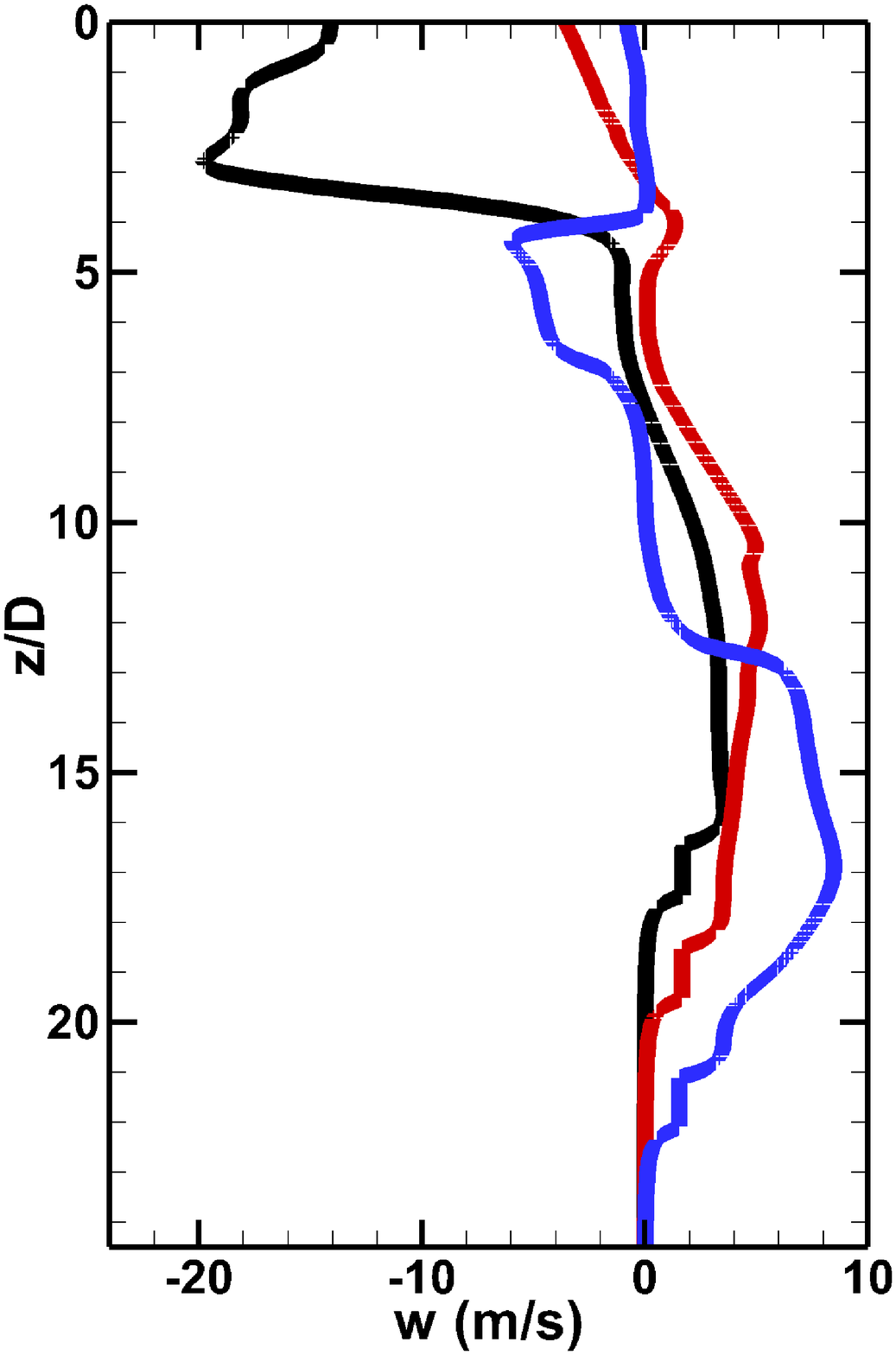}
 \caption{Vertical variation of the (\textit{a}) pressure, and (\textit{b}) vertical velocity, $w$  at $ x,y=0$ line for $W_0=2.3m/s, R_0=1mm$  $(We=72 , Bo= 0.14)$.}
 \label{fig:6} 
\end{figure}

\begin{figure}[ht]
\centering
(a)\includegraphics[width=6cm]{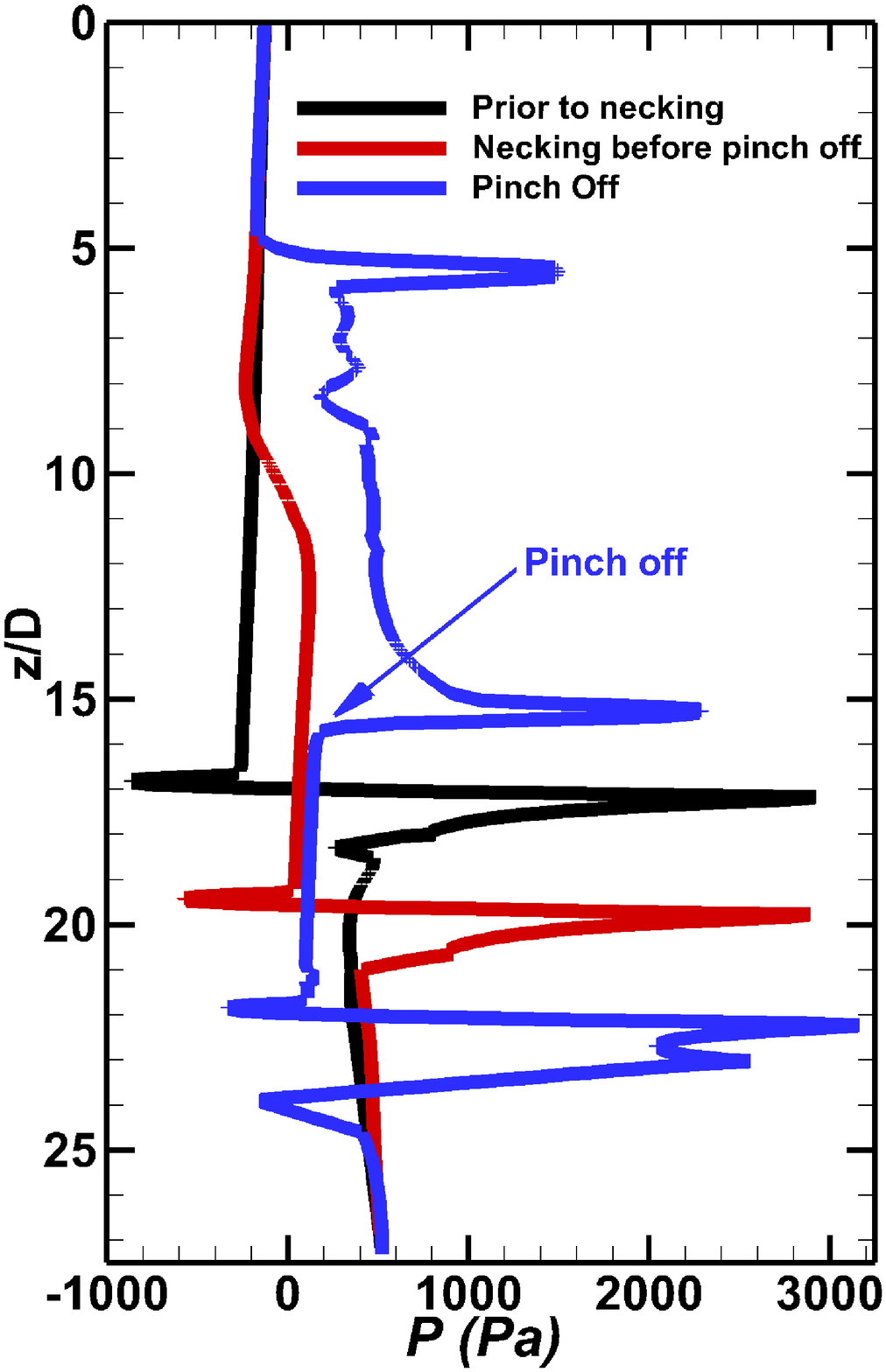}
(b)\includegraphics[width=6cm]{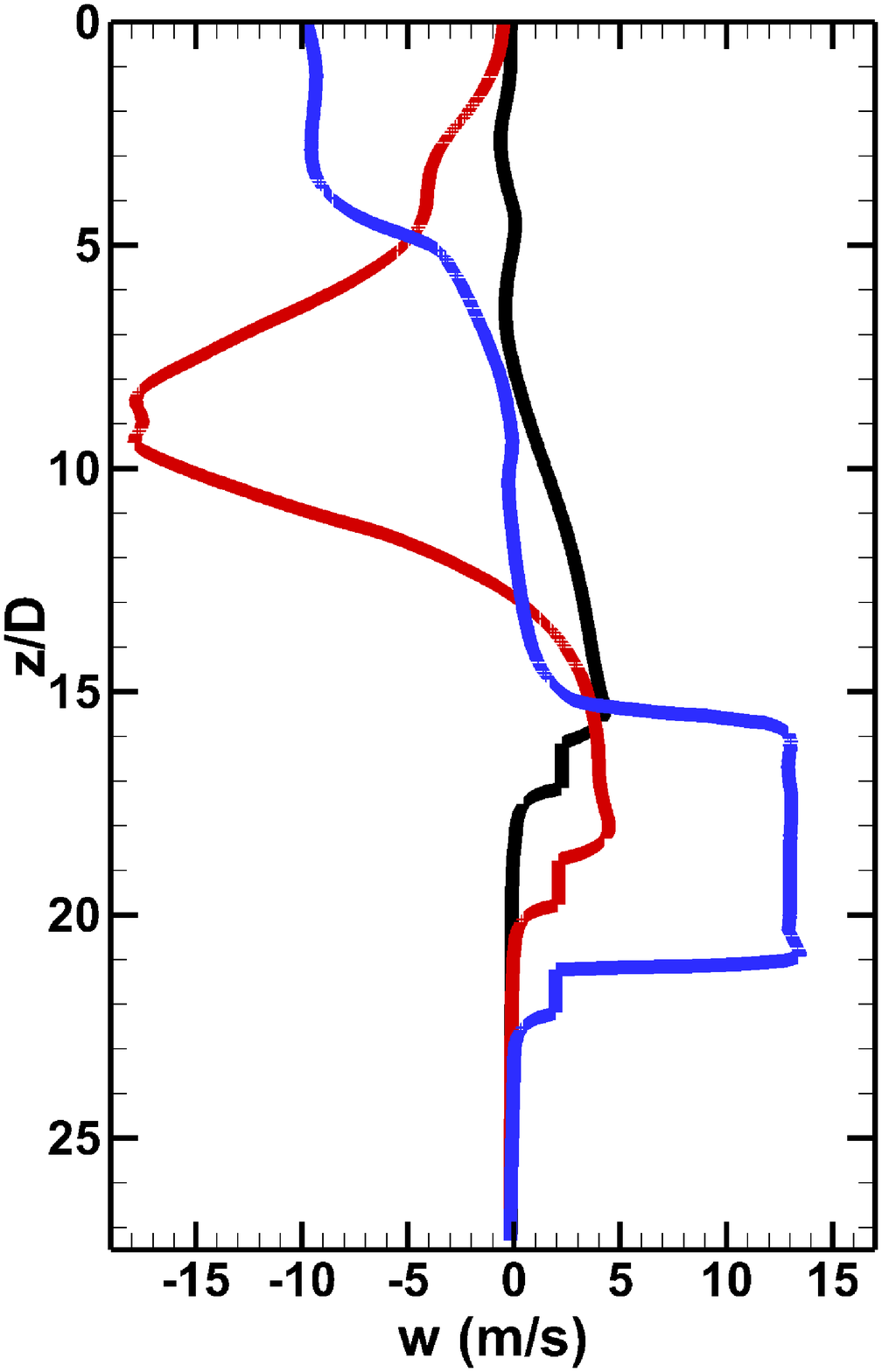}
\caption{Vertical variation of the (\textit{a}) pressure, and (\textit{b}) vertical velocity, $w$ at $ x,y=0$ line for $W_0=3.1m/s, R_0=0.79mm$  $(We=109 , Bo= 0.088)$.}
\label{fig:7}
\end{figure}

\begin{figure}[ht]
 \centering
(a)\includegraphics[width=6cm]{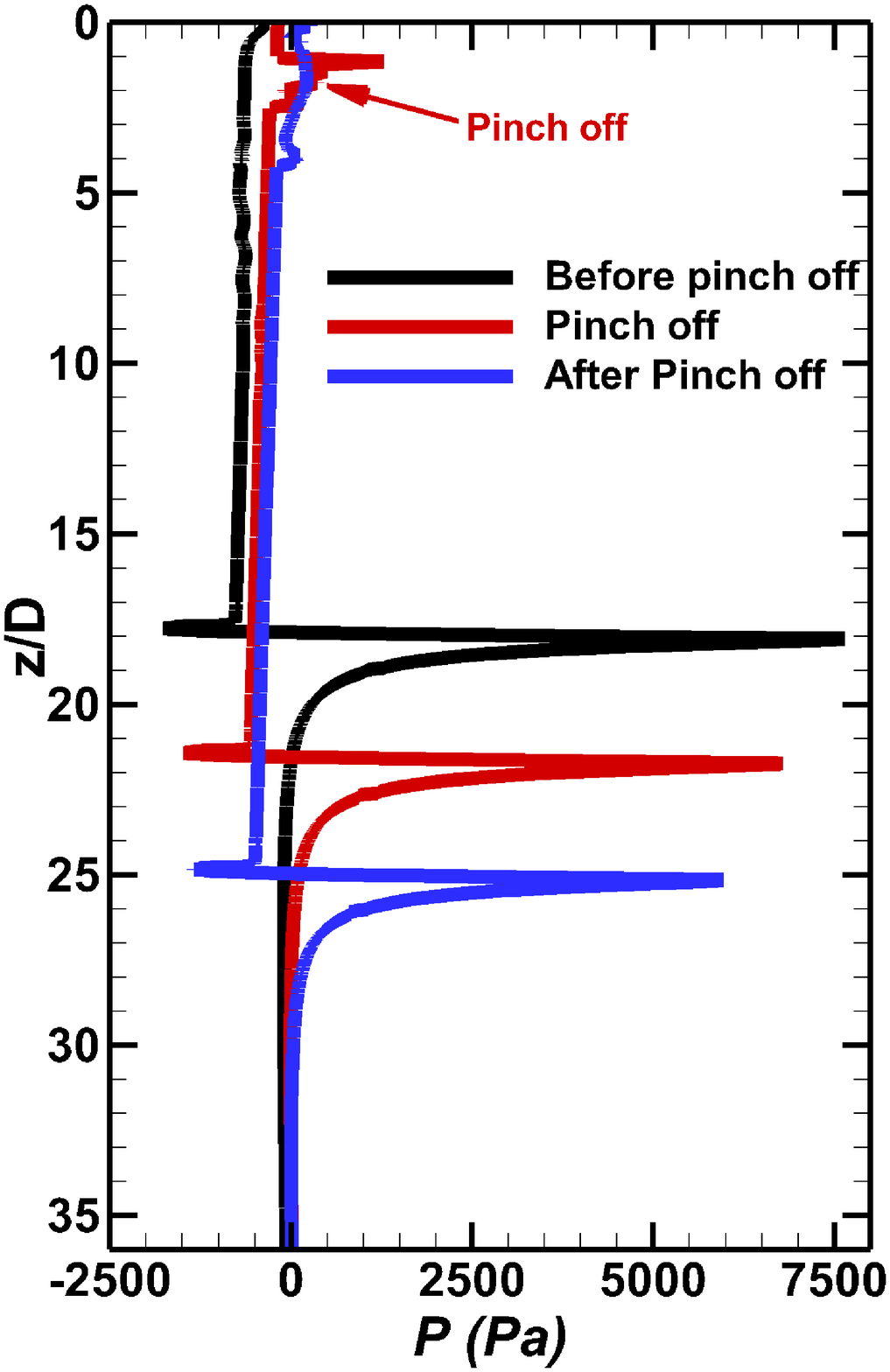}
(b)\includegraphics[width=6cm]{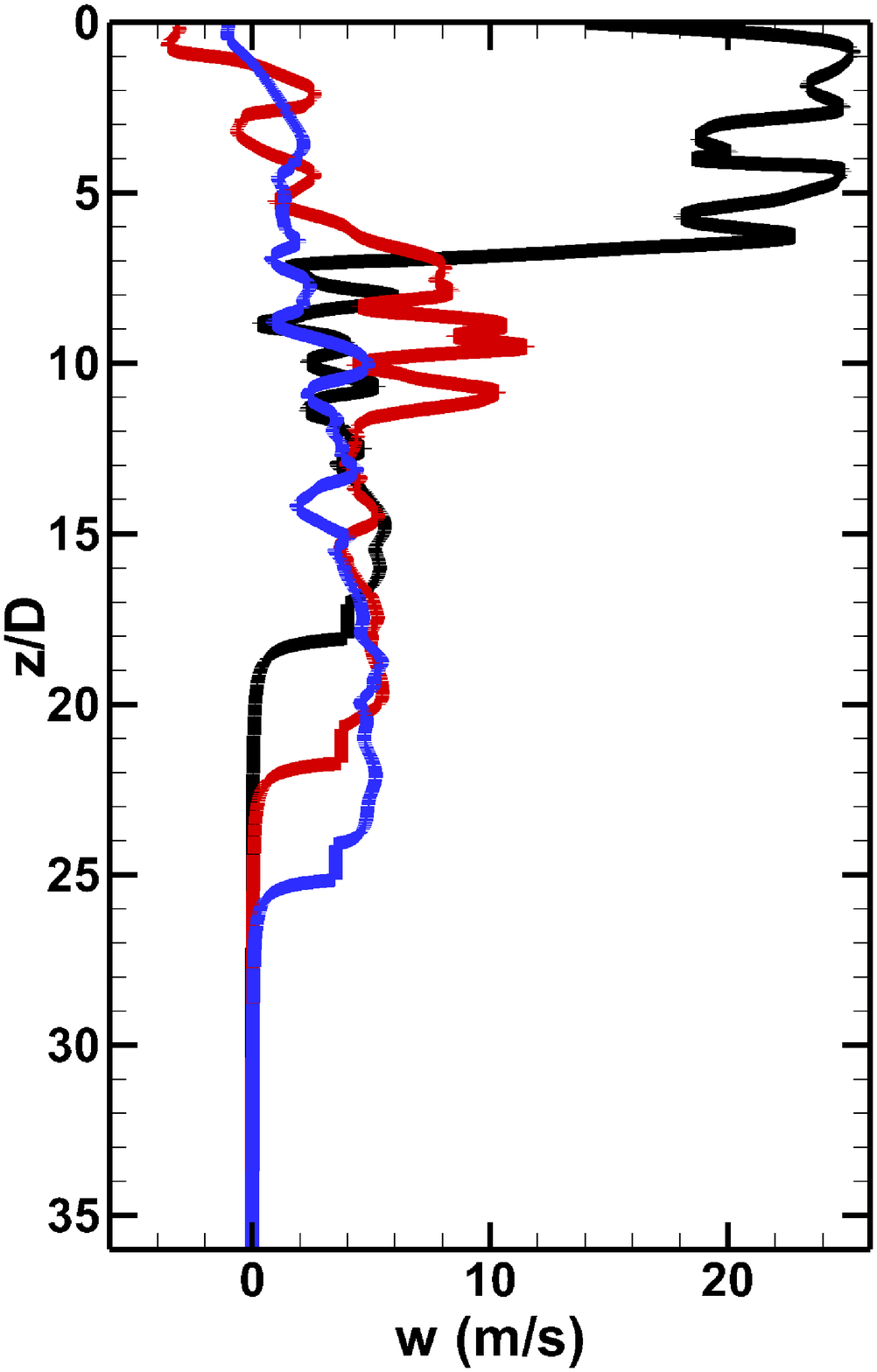}
\caption{Vertical variation of the (\textit{a}) pressure, and (\textit{b}) vertical velocity, $w$ at $ x,y=0$ line for $W_0=5.4 m/s, R_0=1 mm$  $(We=420 , Bo= 0.14)$.}
\label{fig:8} 
\end{figure}

We further assess the cavity dynamics from the vertical variation of the pressure and the vertical velocity field along the $x=0,y=0$ line at different time instances. For $W_0=2.3$m/s, $R_0=1$mm $(We=72, Bo= 0.14)$, the vertical variation of pressure and velocity along $ x,y=0$ line is demonstrated in figure \ref{fig:6}(a) and (b) respectively, at time instances before the pinch-off, at the pinch off and the second pinch off. The maximum pressure is at the bottom of the sphere due to fluid stagnation. A sharp increase in pressure at the pinch-off location also occurs due to the stagnation of the fluid. We demonstrate this by the arrows in figure \ref{fig:6}(a). The pinch-offs take place owing to the dominance of surface tension forces over viscous forces. The vertical velocity is zero at the pinch-off locations. Note that the pressure inside the cavity decreases linearly. This decrease in pressure occurs due to the expansion of the sealed cavity. During the first pinch-off, the velocity becomes negative, signifying air leaving the cavity in the form of a nozzle in the upwards direction. We also observe a negative velocity at the location of the separated air bubble. This indicates the formation of an upward-moving jet within the air bubble, as also demonstrated in the supplementary Movie 1. Notice that the vertical velocity becomes approximately zero after the second pinch-off before becoming negative due to the momentum balance between the downward-moving fluid and the upward-moving jet of air.\\

For $W_0=3.1 m/s, R_0=0.79 mm$ $(We=109, Bo= 0.088)$, the pressure and velocity variation in the vertical direction along $ x,y=0$ line at time intervals: before necking, at necking and pinch off after necking are shown in figures \ref{fig:7} (a) and (b) respectively. In this case, a deep seal pinch-off occurs. The pressure distribution is similar to that observed during the second pinch-off for the previous case. We can see that the pressure inside the cavity decreases linearly due to the expansion of the cavity. During the necking phenomenon, negative velocity indicates an upward flow of air. The supplementary Movie 2 demonstrates this process. At the necking region, the air cavity becomes narrow. The air inside the cavity is at a pressure higher than the surrounding, and tries to escape from this narrow necking region. Since the air is trying to escape from a converging region, the air velocity in the upward direction increases significantly, as indicated by a negative velocity in figure \ref{fig:7}(b). At pinch-off (indicated by the blue arrow in figure \ref{fig:7}(a)), the fluid almost stagnates, increasing pressure in the zone. The vertical velocity remains constant inside the air cavity. However, post pinch-off, the vertical velocity nearly becomes zero, similar to that observed for the previous case.\\

The pressure and the vertical velocity plot at $(We=420, Bo= 0.14)$ for three-time intervals: before the pinch-off, at pinch-off, and after pinch-off; $W_0=5.4m/s, R_0=1mm$ is shown in figures \ref{fig:8} (a) and (b). We can observe that the velocity before pinch-off is highly positive, indicating the rapid downward movement of the air into the cavity formed owing to the fast-moving sphere into the pool of water. The rapid movement of the air results in negative pressure at this location, and therefore the water rushes towards the axis of the cavity, eventually leading to the pinch-off. The pinch-off occurs near the surface as manifested by the increase in the pressure (indicated by the red arrow in figure \ref{fig:8}(a)). We can also observe that the vertical velocity inside the cavity fluctuates and becomes negative at the pinch-off location, signifying the upwards movement of the air. \\
 
We also assess the drag force for the different cases by evaluating the drag coefficient. Additionally, we simulate the translation of a hydrophilic sphere at $Bo, We = (0.14, 132), (0.088, 109)$ and compare its drag coefficient with the analogous hydrophobic case. When a hydrophobic sphere translates through water, the lower half remains in contact with water, and the upper half remains inside the air, thus, resulting in a significant reduction in the drag coefficient.  However, for the hydrophilic sphere, no such cavity forms during its translation through the pool of water.\\


\begin{figure}[h]
\centering
(a) \includegraphics[width=7.5cm]{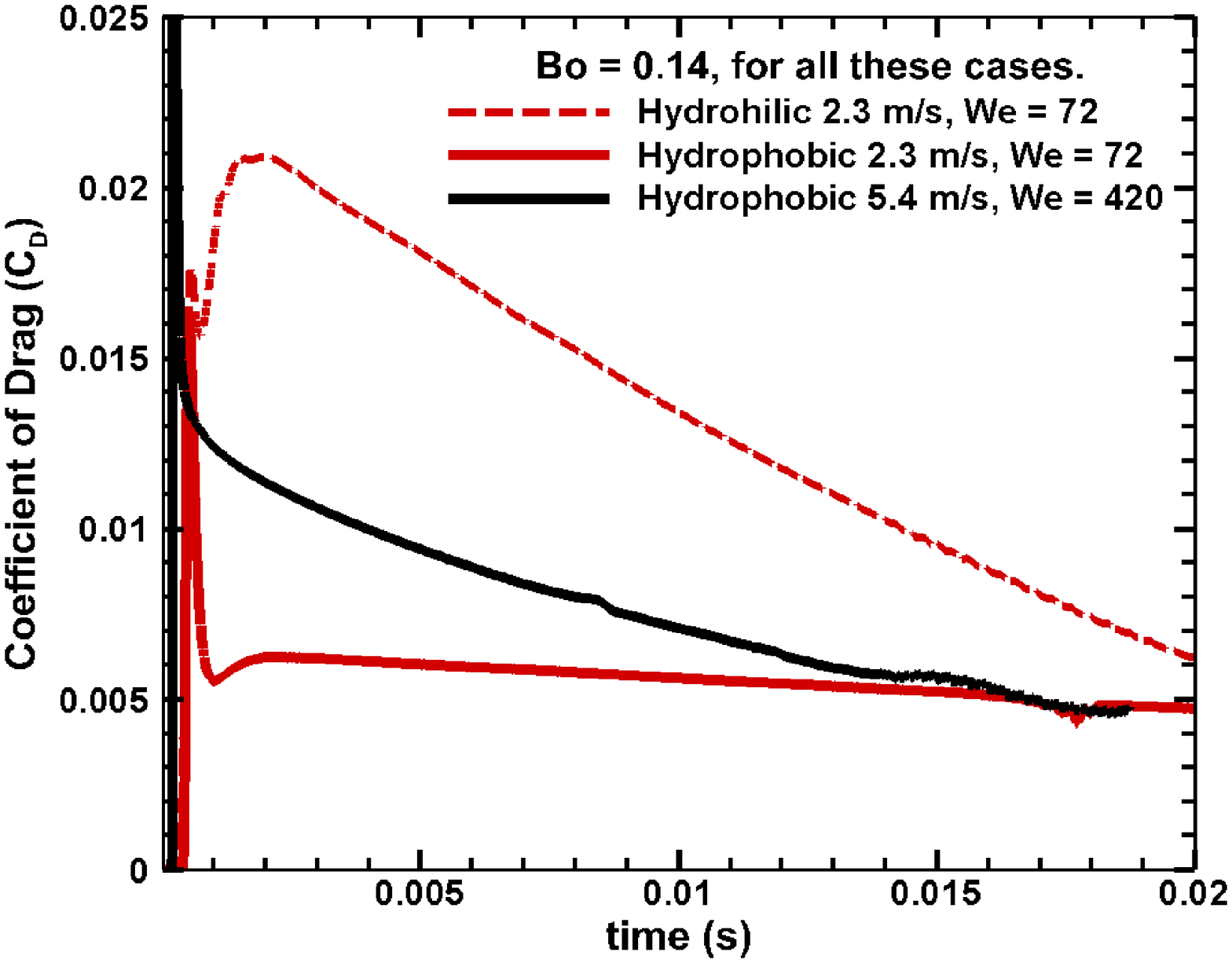}\\
(b) \includegraphics[width=7.5cm]{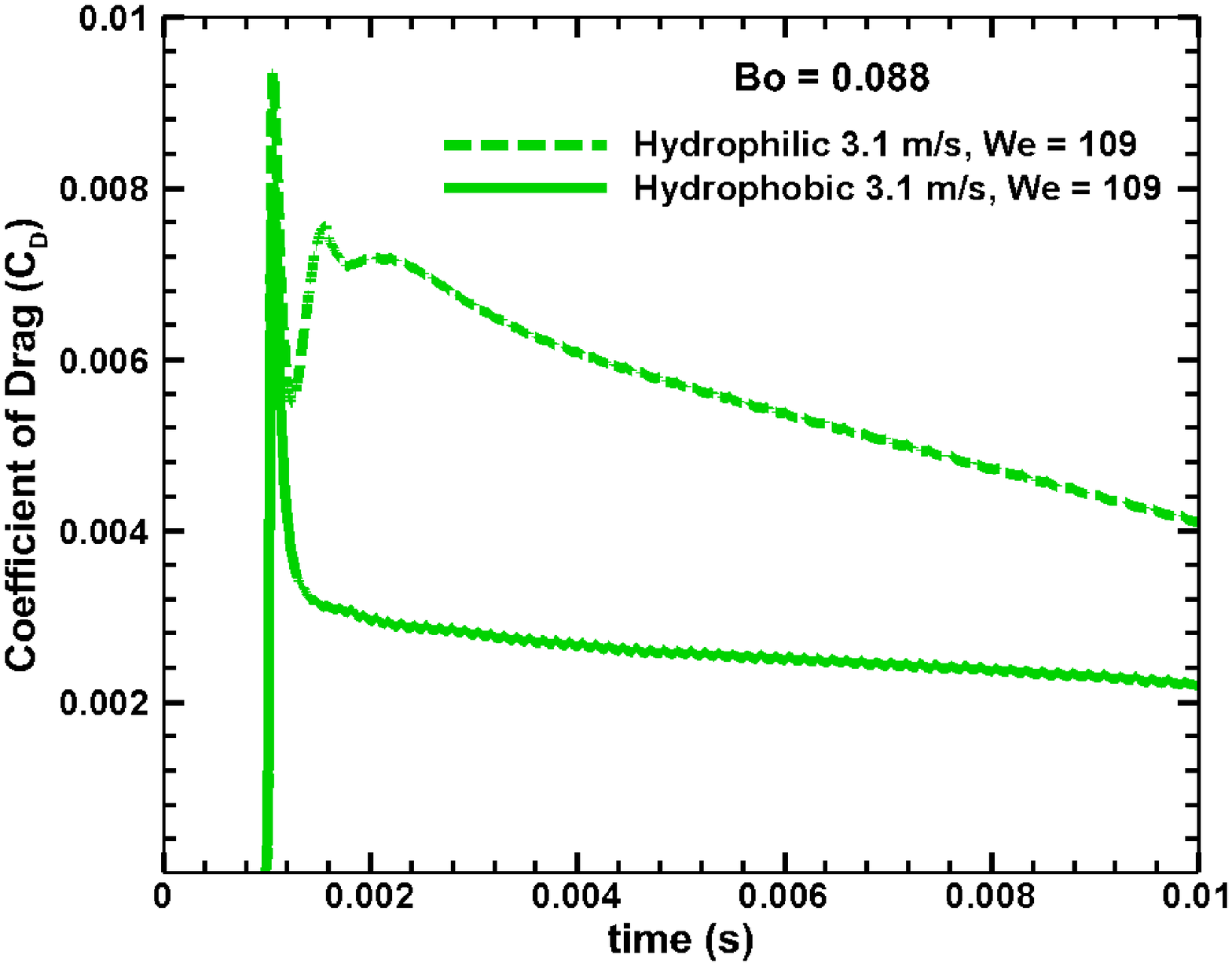}
\centering
\captionof {figure} {Comparison of the coefficient of drag $(C_D)$ (\textit{a}) with velocities $2.3m/s$, $3.1m/s$ and $5.4m/s$ at $Bo=0.14$, (\textit{b}) at velocity $3.1$ m/s and $Bo = 0.088$.}
\label{fig:9}%
\end{figure}

Figure \ref{fig:9}(a) shows the comparison of the coefficient of drag ($C_D$) among different cases with the same $Bo = 0.14$ and different Weber numbers. We have also included a hydrophilic case as $We = 72$ to estimate the drag reduction for a hydrophobic case at the same $We$. Similarly, figure \ref{fig:9}(b) demonstrates the time evolution of $C_D$ for the hydrophilic and hydrophobic spheres at $We = 109$ and $Bo = 0.088$. We measure $C_D$ when the sphere touches the water surface. In all the cases, $C_D$ reaches its peak value when a large portion of the sphere submerges in the water for a short period. The peak of $C_D$ depends on the speed at which the sphere enters water. We find that for a higher impact speed of the sphere, the initial $C_D$ is high. When the sphere completely submerges into the water, $C_D$ starts dropping. At a higher speed of $5.4$ m/s ($We = 420$), the decrease in the drag coefficient is significantly quicker than that at $2.3$ m/s ($We = 72$). We can also observe that the $C_D$ for a hydrophobic sphere is less than the hydrophilic sphere at $We = 72, 109$ and $Bo = 0.14, 0.088$. For the hydrophobic sphere, only the lower portion remains in contact with water, whereas the upper part remains inside the air cavity resulting in lower drag. In contrast, the absence of an air cavity in the hydrophilic spheres results in relatively more drag. \\

\section{Conclusions}
We perform numerical simulations to study the air cavity that forms during the translation of a hydrophobic sphere from air to water at different speeds. When a sphere of radius $1$ mm hits the water surface at a speed of $2.3$ m/s, it produces a tongue-shaped air cavity that pinches off very near the surface. Such cavities are known as shallow seals. A sphere of radius $0.79$ mm with an impact speed of $3.1$ m/s results in a similar air cavity. However, this cavity pinches off at a depth relatively greater than the sphere impacting at $2.3$ m/s and is known as a deep seal cavity. When the sphere with a radius of $1$ mm impacts the pool of water at a speed of $5.4$ m/s, a splash curtain domes over to close the air cavity from above, forming a surface seal cavity. The qualitative features of the different types of cavities obtained from our simulations agree very well with the experiments of \cite{aristoff_bush_2009}. We further report the pressure and the vertical velocity distribution along a vertical line passing through the center of the sphere to assess the cavity dynamics quantitatively. The pressure increases at the locations of the pinch-offs for all the cases due to the stagnation of the fluid at the pinch-off locations. We also capture the fine details associated with the dynamics of air cavities at different speeds. At $2.3$ m/s, we see two pinch-offs, along with the existence of a tiny air bubble and the formation of a jet inside the secondary air bubble separated from the big air cavity. Similarly, for $5.4$ m/s, we observe a splash curtain and a water jet penetrating the air cavity from above. All these details agree with the experiments of \cite{aristoff_bush_2009}.\\ 

We also present a comparison of the drag coefficient among the different cases. When the sphere comes in contact with the water, we observe a peak in the value of $C_D$ for all the cases. As the sphere submerges in the pool of water, an envelope of air surrounds it, resulting in a decrease in the drag coefficient. We find that with an increase in the impact speed from $2.3$ m/s to $5.4$ m/s, the rate of decrease of $C_D$ increases. We also compare the drag coefficient between hydrophobic and hydrophilic spheres and find that the presence of the air cavity behind the hydrophobic sphere significantly reduces the drag.\\

\bibliographystyle{cas-model2-names}
\bibliography{cas-refs}

\end{document}